\documentclass[aps,twocolumn]{revtex4}
\usepackage{graphicx}
\usepackage{latexsym}
\usepackage{changes}
\begin{document}

\title{Unconventional superconductivity induced by suppressing an iron-selenium based Mott insulator CsFe$_{4-x}$Se$_4$}
\author{Jin Si, Guan-Yu Chen, Qing Li, Xiyu Zhu, Huan Yang and Hai-Hu Wen}\email{hhwen@nju.edu.cn}

\affiliation{National Laboratory of Solid State Microstructures and Department of Physics, Center for Superconducting Physics and Materials, Collaborative Innovation Center for Advanced Microstructures, Nanjing University, Nanjing 210093, China}

\date{\today}

\begin{abstract}
There are several FeSe based superconductors, including the bulk FeSe, monolayer FeSe thin film, intercalated K$_x$Fe$_{2-y}$Se$_2$ and Li$_{1-x}$Fe$_x$OHFeSe, etc. Their normal states all show metallic behavior. The key player here is the FeSe layer which exhibits the highest superconducting transition temperature in the form of monolayer thin film. Recently a new FeSe based compound, CsFe$_{4-x}$Se$_4$ with the space group of $Bmmm$ was found. Interestingly the system shows a strong insulator-like behavior although it shares the same FeSe planes as other relatives. Density functional theory calculations indicate that it should be a metal, in sharp contrast with the experimental observations. Here we report the emergence of unconventional superconductivity by applying pressure to suppress this insulator-like behavior. At ambient pressure, the insulator-like behavior cannot be modeled as a band insulator, but can be described by the variable-range-hopping model for correlated systems. Furthermore, the specific heat down to 400 mK has been measured and a significant residual coefficient $\gamma$$_0$=$C/T$$\vert$$_T$$_\rightarrow$$_0$ is observed, which contrasts the insulator-like state and suggests some quantum freedom of spin dynamics. By applying pressure the insulator-like behavior is gradually suppressed and the system becomes a metal, finally superconductivity is achieved at about 5.1 K. The superconducting transition strongly depends on magnetic field and applied current, indicating a fragile superfluid density. Our results suggest that the superconductivity is established by diluted Cooper pairs on top of a strong correlation background in CsFe$_{4-x}$Se$_4$.
\end{abstract}

\maketitle

\section{Introduction}

Being different from the iron arsenic based superconductors, the iron selenium (FeSe) based ones have intriguing properties and attract a lot of attentions\cite{44,43,2,3,4,5,6,7,40,8,9,10,11,12,13,14,41,15,1}. The FeSe has the simplest structure and the critical temperature ($T$$_c$) of the bulk samples\cite{44} can be enhanced from about 8 K to 37 K by pressure\cite{43,2,3}. The Meissner shielding signal was observed up to 65 K for FeSe monolayer thin film grown on Nb-doped SrTiO$_3$ substrate\cite{4,5,6}, which could be the highest $T$$_c$ of all IBSs discovered so far (in terms of Meissner effect). By intercalating alkali metals between FeSe layers, superconductivity was observed in a series of compounds, such as A$_x$Fe$_{2-y}$Se$_2$ (A = Na, K, Rb, Cs, Tl/Rb, and Tl/K)\cite{40,7,8,9,10,11,12,13} and Li$_{1-x}$Fe$_x$OHFeSe\cite{14}, and $T$$_c$ can be enhanced to 32 K, or even to 46 K. Also, the recently found hole type FeSe based superconductor S$_{0.24}$(NH$_{3}$)$_{0.26}$Fe$_2$Se$_2$ enriches the physics in the family of FeSe based superconductors\cite{41}. An obvious common feature for all of them is that they have the FeSe layers as the conducting sheets from which the superconductivity originates. The normal states of these systems show clear metallic behavior. Recently, a new FeSe based compound CsFe$_{4-x}$Se$_4$ was found\cite{15}, it shows distinct features although it has the roughly perfect FeSe layer. Unlike the system A$_x$Fe$_{2-y}$Se$_2$ which has a clear phase separation, the CsFe$_{4-x}$Se$_4$ is uniform without the trace of phase separation. This compound shows an insulator-like behavior and does not undergo clear antiferromagnetic transitions. At room temperature, it forms an orthorhombic lattice structure with a space group of $Bmmm$, however bulk FeSe undergoes a structural transition from tetragonal to orthorhombic at around 90 K\cite{1,43,2}. The density functional theory (DFT) calculations on CsFe$_{4-x}$Se$_4$ indicate that it should be a metal with an intermediate density of states (DOS)\cite{15} at the Fermi level comparing with the value of FeSe and A$_x$Fe$_{2-y}$Se$_2$, which is in sharp contrast with the experimentally observed insulator-like behavior. The insulator-like behavior in CsFe$_{4-x}$Se$_4$ remains puzzling and elusive.

In this paper, we report the successful synthesis of this new compound with high quality and measurements of its intriguing features. At ambient pressure, it behaves as an insulator, and the resistivity obeys the relation $\ln$$\rho$$\propto$(1/$T$)$^{1/4}$ in a wide temperature range, which is consistent with the prediction of variable-range-hopping (VRH) model for correlated systems\cite{16}. In the doped Mott system, the effective charges, like holes in underdoped cuprate, move on an inhomogeneous background with strong electronic correlation, exhibiting the VRH behavior of motion\cite{39}. When measuring the specific heat down to 400 mK, a sizable residual coefficient can be observed, which is in sharp contrast to the strong insulator-like behavior of the system. By applying pressure, a transition from insulator to metal occurs and also a superconducting transition appears at about $T$$_c$ = 5.1 K. The zero resistance temperature depends strongly on external magnetic field and applied current, suggesting a fragile superfluid density. Our results indicate that the superconductivity is established by inducing diluted Cooper pairs on top of the FeSe planes which exhibits a background of strong correlation.

\section{Methods}

By using solid state reaction method, polycrystalline samples of CsFe$_{4-x}$Se$_4$ were successfully synthesized. Firstly, FeSe precursors were prepared by reacting Fe powder and Se powder (Alfa Aesar, 99.999${\%}$, and 99.99${\%}$, respectively) at 950 K for 24 hours in sealed quartz tubes. Secondly, we reground the FeSe precursors, mixed them with stoichiometric amount of alkali metal Cs (Alfa Aesar, 99.95${\%}$) in an alumina crucible and sealed the crucible into an evacuated quartz tube under vacuum. All manipulations were performed in a glove box filled with argon gas. The tube was subsequently heated up to and kept at 873 K for 24 h. Then, the obtained sample was pulverized, pressed into a pellet, sealed in a quartz tube under vacuum, heated up and kept at 923 K for 72 hours. Finally, samples with dark color were obtained. The sample is sensitive to air and thus usually it is vacuum-packed and put in the glove box. The synthesis process is similar to that of previous work\cite{15}. The only difference is that in previous work, the quartz tube was sealed with Ar gas, while in our experiments, the quartz tube was sealed under evacuation.

The X-ray-diffraction (XRD) data were obtained with a Bruker D8 Advanced diffractometer and the Rietveld refinements\cite{17} were conducted with the TOPAS 4.2 software\cite{18}. The energy dispersive spectrums (EDS) measurements were carried out with a Phenom ProX scanning electron microscope (SEM) at an accelerating voltage of 15 kV. The resistivity data at ambient pressure was measured on a Physical Property Measurement System (PPMS, Quantum Design). Gold wires with diameter of 30 micrometers were attached to the sample with silver paint forming a standard four-probe configuration and the contact resistance is in magnitude of 1 $\Omega$. The magnetization measurements were performed on a SQUID-VSM (Quantum Design). The specific heat was measured with thermal-relaxation method by an option of the PPMS with a He$^3$ insert which allows us to measure specific heat down to 0.4 K. The resistivity data under high pressure were obtained by using a diamond-anvil-cell (DAC) module (cryoDAC-PPMS, Almax easyLab). The Pt electrodes were attached to the sample with a four-probe van der Pauw method\cite{19} and the contact resistance is in magnitude of 1 $\Omega$ above 1 GPa. The pressure medium used in the DAC is fine powder of NaCl and the gasket was made of T301 stainless steel. The applied pressures were calculated by measuring the shift of ruby R$_1$ luminescent line\cite{20}.

\section{Results}

\begin{figure}
  \includegraphics[width=7.5cm]{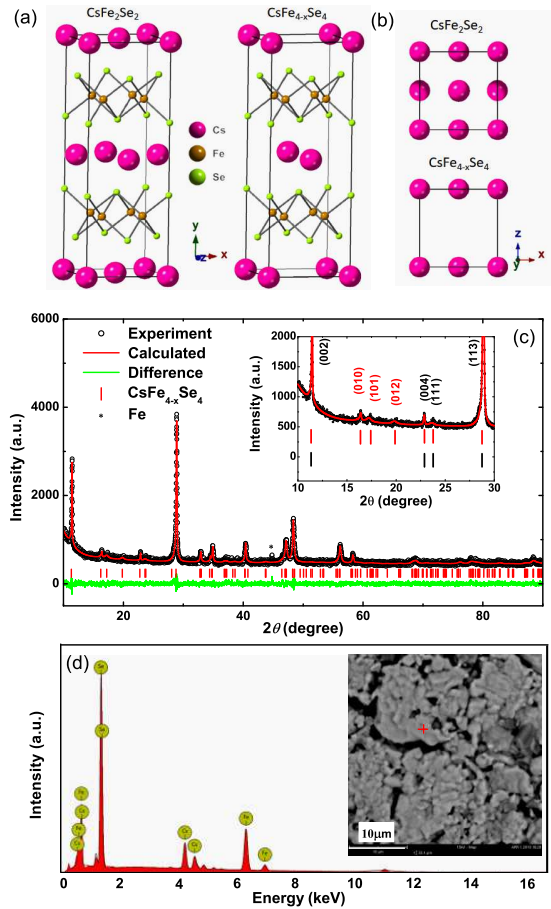}
\caption{Structure and compositional analysis of CsFe$_{4-x}$Se$_4$. (a), Schematic crystal structure of CsFe$_2$Se$_2$ and CsFe$_{4-x}$Se$_4$. (b), Crystal structure of CsFe$_2$Se$_2$ (restructured) and CsFe$_{4-x}$Se$_4$ viewed along [001] direction. (c), The x-ray-diffraction pattern for polycrystalline CsFe$_{4-x}$Se$_4$ with the Rietveld refinement. Inset in (c), Typical peaks of CsFe$_{4-x}$Se$_4$ are marked with red indices. The black vertical lines show the coincident indices of CsFe$_2$Se$_2$ and CsFe$_{4-x}$Se$_4$. (d), A typical energy dispersive spectrum which shows the compositional ratio of the measured spot. The averaged composition of Cs:Fe:Se = 1.2:3.95:4 was obtained by the measurements at 19 spots and areas. The inset shows a scanning electron micrograph image.} \label{fig1}
\end{figure}

Comparing with the close related structure Cs$_{1-x}$Fe$_{2-y}$Se$_2$ (for simplicity we denote it as CsFe$_2$Se$_2$ hereafter), the new compound CsFe$_{4-x}$Se$_4$ naturally possesses the orthorhombic symmetry at room temperature. Schematic structures of these two compounds are shown in Fig.~\ref{fig1}(a). The crystal structure of CsFe$_2$Se$_2$ is also plotted in order to have a close comparison with that of CsFe$_{4-x}$Se$_4$. The Cs plane in the middle of the structure of CsFe$_2$Se$_2$ is replaced by the Cs one-dimensional chains in CsFe$_{4-x}$Se$_4$, so that CsFe$_{4-x}$Se$_4$ possesses the C2 symmetry instead of the C4 symmetry, this difference is shown directly in Fig.~\ref{fig1}(b) as the top views of the Cs structures for the two systems. The X-ray-diffraction (XRD) pattern with the Rietveld refinement\cite{17} (Fig.~\ref{fig1}(c)) and the scanning electron micrograph (SEM) images with energy dispersive spectroscopy (EDS) (Fig.~\ref{fig1}(d)) are measured to verify the phase of our sample. The Rietveld refinements are carried out with the TOPAS 4.2 software\cite{18}. From the refinements, we can obtain the agreement factors $R$$_{wp}$ = 4.69${\%}$ and $R$$_{p}$ = 3.70${\%}$ and the relatively small values mean that the calculated profile agrees with our experimental data quite well. The lattice parameters determined here are $a$ = 5.45(2) ${\mathring{A}}$, $b$ = 5.46(0) ${\mathring{A}}$, $c$ = 15.62(8) ${\mathring{A}}$, which are close to those of previous work\cite{15}. The inset gives the details of the XRD data in the low-angle region, which shows the typical peaks owing to the C2 symmetry. The phase indices marked with black vertical lines are calculated by the CsFe$_2$Se$_2$ phase with the converted lattice parameter ($a$ = $b$ = 3.852 ${\mathring{A}}$, $c$ = 15.628 ${\mathring{A}}$), while the red indices belong to CsFe$_{4-x}$Se$_4$. Here the black vertical lines show the coincident positions of CsFe$_2$Se$_2$ and CsFe$_{4-x}$Se$_4$, which does not mean the existence of CsFe$_2$Se$_2$. A tiny peak at about 45 degrees comes from the impurity of iron and the ratio is about 0.6${\%}$ in atomic ratio, which has been marked in Fig.~\ref{fig1}(c). So that our sample is of good quality and does not contain the impurity phase like precursors FeSe. We determine the occupancies of Fe and Cs by doing the Rietveld refinement of the XRD data and the obtained occupancies of Fe and Cs are 0.908 and 1.045, respectively. Meanwhile, we use the x-ray energy dispersive spectrum to analyze the compositions of Cs, Fe and Se in the samples (for details please refer to Methods). The mean value of the ratio of the three elements is roughly Cs:Fe:Se = 1.2:3.95:4, which means our samples have a little excess Cs and small amounts of Fe vacancies comparing with the stoichiometric standard formula. The existence of Fe vacancy seems to be a common feature in the intercalated FeSe systems\cite{8,9,11}, and the excess Cs may exist in the form of Cs oxides because Cs is very active, and a weak reaction with slight amount of residual oxygen may be inevitable during the synthesis process.

\begin{figure}
  \includegraphics[width=8.5cm]{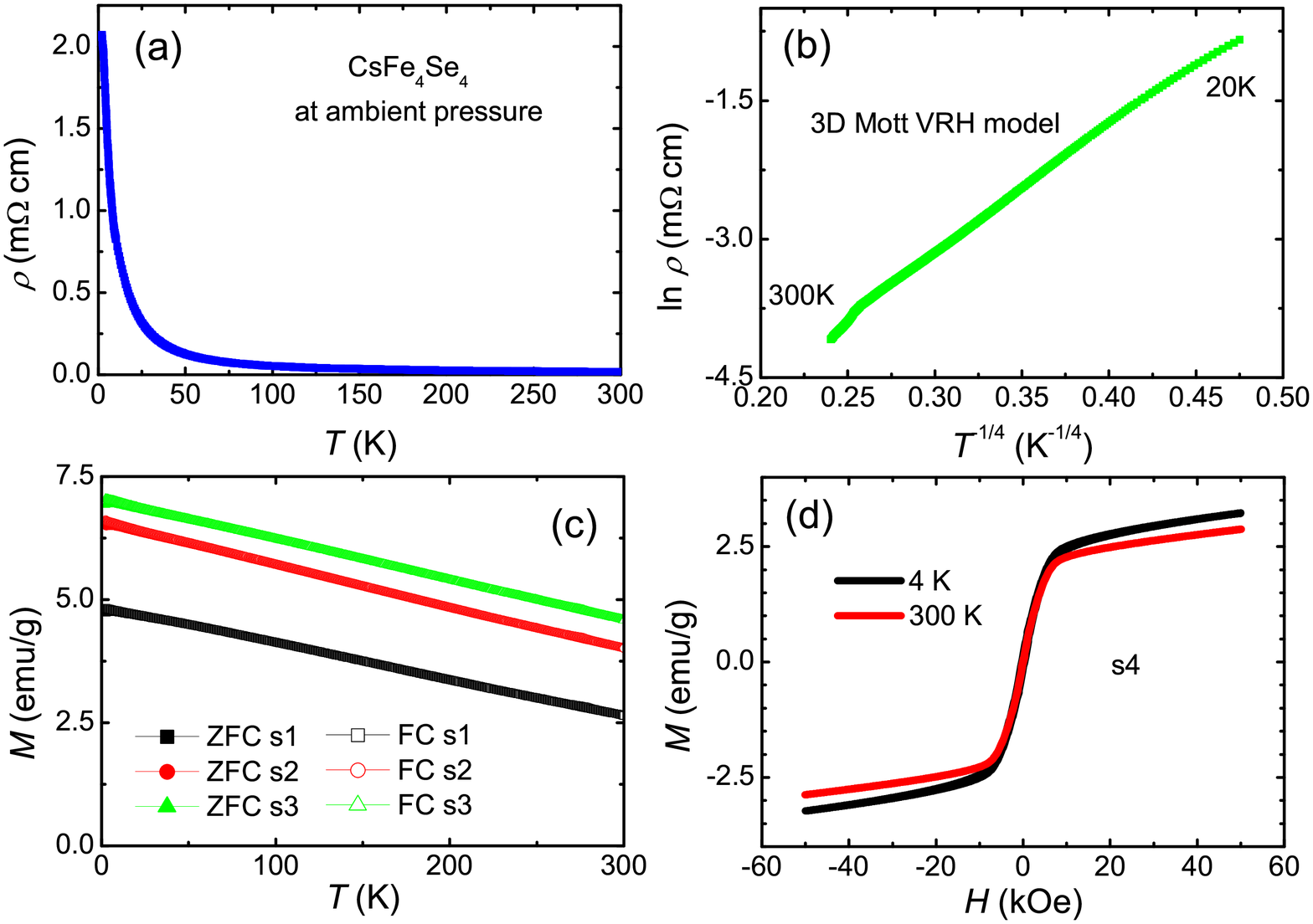}
\caption{Resistivity and magnetization of CsFe$_{4-x}$Se$_4$. (a), The temperature dependence of resistivity of CsFe$_{4-x}$Se$_4$ at ambient pressure. (b), Resistivity data from 20 K to 300 K show a rough linear behavior, which is consistent with the variable-range-hopping (VRH) model for the three-dimensional correlated systems. (c), Temperature dependence of magnetization measured at the field of 1 T for three samples. (d), Magnetic field dependence of magnetization measured at 4 K and 300 K for another sample.} \label{fig2}
\end{figure}

As mentioned before, the resistivity of CsFe$_{4-x}$Se$_4$ polycrystalline sample behaves as an insulator, which is shown in Fig.~\ref{fig2}(a). As shown in Fig.~\ref{fig2}(b), the data between 20 K to 300 K are fitted roughly by the formula of three-dimensional variable-range-hopping (3D VRH)\cite{16}, namely $\rho$=$\rho$$_{0}$$\cdot$exp[($T$$_{0}$/$T$)$^{1/4}$]. To our surprise is that this relation roughly holds in wide temperature region. The slight deviation below 20 K may indicate the possible change of the conduction mechanism or modification to the 3D VRH model. We also try to fit the resistivity data with other possible models, such as the band gap model ($\rho$=$\rho$$_{0}$$\cdot$exp($T$$_{0}$/$T$)), 1D ($\rho$=$\rho$$_{0}$$\cdot$exp[($T$$_{0}$/$T$)$^{1/2}$]) and 2D ($\rho$=$\rho$$_{0}$$\cdot$exp[($T$$_{0}$/$T$)$^{1/3}$]) VRH models\cite{16,21} and the small polaron hopping model\cite{22} ($\rho$=$\rho$$_{0}$$T$$\cdot$exp($T$$_{0}$/$T$)), but all fitting are failed. One can see these model fittings in in Supplementary Figure 1. Indeed, insulating grain boundaries in granular samples would also contribute such an insulator-like feature. However, we did not see such secondary insulating phase (if they would exist on the surface of grains) from XRD. If this insulator-like behavior arises just from the scattering of the grain boundaries (supposing metallic grains), under a high pressure, such as several GPa, we would not believe that the electric conduction through the grain boundaries is still insulator-like.

The temperature dependences of magnetization measured for three samples at the magnetic field of 1 T in both zero-field-cooling (ZFC) and field-cooling (FC) modes are plotted in Fig.~\ref{fig2}(c). It is easy to find that the magnetization follows a rough linear decreasing behavior with temperature in wide temperature region, this certainly violates the Curie-Weiss law and the reason for the strange linear temperature dependence of magnetization is still unknown. The difference of magnetization values (about 25${\%}$) between different samples in Fig.~\ref{fig2}(c) may be induced by the different content of Fe impurities. Although the amount of Fe impurity is tiny, due to its ferromagnetic feature of Fe, this will lead to different background of total signal of magnetization. The content of Fe impurity in different samples should be slightly different so that there would be a difference of magnetization values. A slight difference (about 2${\%}$ at 2 K) between the magnetizations measured in the ZFC and FC modes below about 20 K is observed. The reason for this small difference is still unknown, and could be attributed to the tiny impurity phase with magnetic hysteresis. The small amount of possible impurity phase cannot be distinguished from the XRD data. If this possible minority phase arises from the residual CsFe$_2$Se$_2$, according to the difference of the ZFC-FC magnetizations, we can roughly estimate the ratio of the CsFe$_2$Se$_2$, which is less than 1 wt${\%}$. We do not see any resistivity and magnetization drops in the low temperature region on the insulating background at ambient pressure, this may also exclude the presence of CsFe$_2$Se$_2$ and FeSe, since otherwise the resistivity and magnetization would show associated drops at corresponding temperatures. Also, we measure the low-temperature M(T) curves at H = 10 Oe with ZFC and FC modes. The data are shown in Supplementary Figure 2 which confirms the absence of superconducting phase in as-prepared CsFe$_{4-x}$Se$_4$ at ambient pressure. From the M(H) curves in Fig.~\ref{fig2}(d), we can see a weak ferromagnetic behavior. We regard this signal weak, because the magnetic susceptibility with external field of 1 T at 4 K is only about 3.4$\times$10$^{-3}$ emu cm$^{-3}$ Oe$^{-1}$, which is much smaller than that in normal ferromagnetic materials (about 1 - 10$^{5}$ emu cm$^{-3}$ Oe$^{-1}$). Thus it is reasonable to attribute this weak ferromagnetic signal to the Fe impurity, which consists also with the XRD data and the values of magnetization in M(T) curves shown in Fig.~\ref{fig2}(c).

From our data, we may exclude the antiferromagnetic or ferromagnetic order below 300 K, since we did not see any transition on the temperature dependence of magnetic susceptibility. From the magnitude of magnetization under 1 T (at least 100 times smaller than that in normal ferromagnetic materials), and the near coincidence of the ZFC and FC magnetizations, we can rule out the presence of ferromagnetic order above 300 K. The coincidence of FC and ZFC magnetizations may also allow us to exclude the glassy magnetism. We could not exclude the existence of antiferromagnetic order above 300 K since our measurement was only done up to 300 K. However, according to the data in previous work\cite{15}, the magnetic susceptibility does not show any peak/kink feature below 800 K, so that the antiferromagnetic order at high temperatures may also be excluded.

\begin{figure}
  \includegraphics[width=8.5cm]{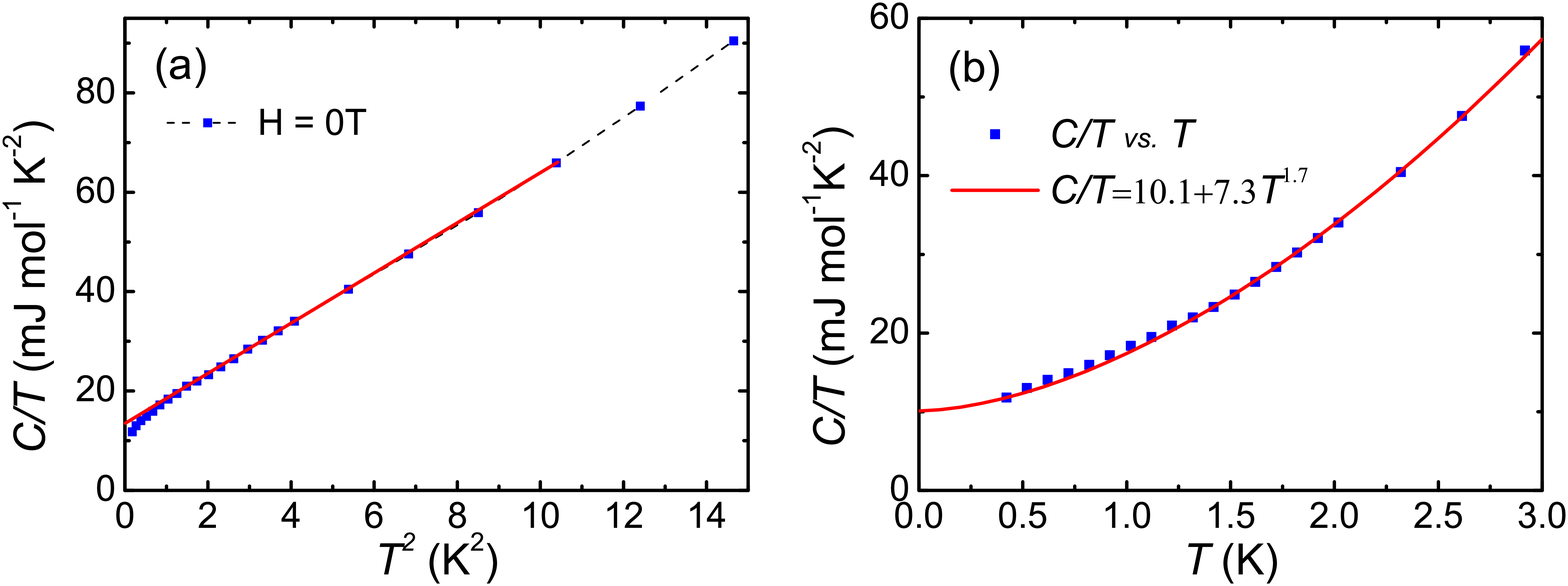}
\caption{Low temperature specific heat of CsFe$_{4-x}$Se$_4$. (a), Quadratic temperature dependence of the specific heat coefficient $C/T$ measured down to 0.4 K. The linear red curve gives the Debye model fit in the temperature range of 1.5 to 4 K, which yields a residual coefficient about 13.6 mJ/mol-K$^2$. (b), Temperature dependence of specific heat coefficient $C/T$ in temperature region from 0.4 K to 3 K. The red curve gives a fitting with a formula $C/T$=10.1+7.3$T$$^{1.7}$ (mJ/mol-K$^2$). A sizable residual specific heat coefficient can be observed.} \label{fig3}
\end{figure}

In order to unravel the puzzling insulator-like behavior of CsFe$_{4-x}$Se$_4$, we have carried out specific heat measurements. Considering the contribution of phonons and conducting electrons in a metal, the specific heat at low temperatures can be described by the Debye model, namely $C/T$=$\gamma$$_0$+$\beta$$T$$^2$+$\delta$$T$$^4$+... Here $\gamma$$_0$ is the specific heat coefficient, $\beta$ and $\delta$ are temperature independent fitting parameters. There should be no residual term $\gamma$$_0$ in above description in the zero temperature limit for a band insulator, since this Somerfield term $\gamma$$_0$ reflects the finite quasiparticle density of states (DOS) at the Fermi energy, which is zero in the band insulator. The terms with higher powers of temperature ($\delta$$T$$^4$+...) should be negligible in the low temperature region, since in the formula of Debye model for the phonon contribution to specific heat, the temperature is normalized by the Debye temperature $\Theta$$_D$ which takes the value of about 147.8 K (see below). In Fig.~\ref{fig3} we present the specific heat data of CsFe$_{4-x}$Se$_4$ down to 0.4 K. Fig.~\ref{fig3}(a) shows the data $C/T$ versus $T$$^2$. One can see two anomalous and intriguing features here. Firstly, the data do not satisfy the simple description of the Debye model. At temperatures above about 1.5 K, one can see a roughly linear behavior of $C/T$ versus $T$$^2$. The red curve shows the Debye model fitting which yields $\beta$ = 5.4 mJ/(mol-K$^4$). According to the Debye model, $\Theta$$_D$=(12$\pi$$^4$$k$$_B$$N$$_A$$Z$/5$\beta$)$^{1/3}$, where $N$$_A$ = 6.02$\times$10$^{23}$ /mol is the Avogadro constant, $Z$ is the number of atoms in one unit cell, here $Z$ = 9 for CsFe$_{4-x}$Se$_4$. Using the obtained value of $\beta$, we get the Debye temperature $\Theta$$_D$ $\approx$ 147.8 K. Below about 1.5 K, however, the curve of $C/T$ vs. $T$$^2$ shows a slight bending down, which is clearly deviating from the description of the Debye model. Secondly, to our surprise, there is a residual term of specific heat coefficient $\gamma$$_0$ in the zero temperature limit concerning the strong insulator-like state. If we follow the linear extrapolation of the Debye model, as highlighted by the red linear line in Fig.~\ref{fig3}(a), we can get a residual term $\gamma$$_0$ = 13.6 mJ/mol-K$^2$ at $T$ = 0 K. This is totally unexpected for a band insulator. Even following the bending down trend of $C/T$ in the low temperature region, we can still see a sizable value of $\gamma$$_0$. In order to carry out the empirical relation of $C/T$ vs. $T$ in low temperature region, we show the raw data in the temperature region of 0.4 to 3 K in Fig.~\ref{fig3}(b), and fit it with an empirical relation $C/T$=$\gamma$$_0$+$a$$T$$^{n}$. The best fitting yields a relation $C/T$=10.1+7.3$T$$^{1.7}$ mJ/mol-K$^2$. Thus a residual term $\gamma$$_0$ clearly exists. Concerning the strong insulator-like behavior seen from the resistivity, this residual term of specific heat in the zero temperature limit indicates a non-trivial origin. Since the residual term $\gamma$$_0$ is quite large, we cannot attribute it to any possible metallic impurities.

In some amorphous compounds, a linear specific heat term could be observed\cite{23}, however, as shown by the XRD data of our sample, the dominant phase here is CsFe$_{4-x}$Se$_4$ with well-formed crystalline structure which does not show any trace of amorphous. In previous studies on the spin glass state of CuMn alloys\cite{24}, a linear term of specific heat was also observed together with a deviation from the Debye model, this was attributed to the extra hyperfine entropy contribution of spins in the spin glass state. In a more general point of view, it was proposed that the linear term of specific heat might exist for a disordered insulator or glassy state, as discussed by Anderson et al.\cite{45} and Phillip\cite{46}. This possibility can be also ruled out since no evidence of spin glass transition is observed in our samples. If there were a spin glass transition, there should be a difference between the ZFC and FC magnetizations below the transition temperature, and a peak of magnetic susceptibility would appear, but all these are not observed in our samples. Although we have slight amount of Fe vacancies and excess Cs in the samples, but they exist not as segregations, in this case they should not contribute a linear term of specific heat. Furthermore, even these Fe vacancies and excess Cs are in disordered state in the sample, due to the very small amount, they should not contribute such a large residual linear term of specific heat. One fact to corroborate this point is that the contents of Fe vacancies and excess Cs are different in our samples and that of Ref.18, but the linear terms of specific heat are quite similar in magnitude. Thus we believe the existence of the linear term of specific heat is intrinsic for the system.

In some spin liquid candidates, a residual term of specific heat exists, which is considered as the contributions of the quantum spin fluctuations of a spin liquid at zero temperature\cite{25}. Thus we intend to conclude that the linear term together with a deviation from the Debye model may be attributed to the nontrivial ground state of the CsFe$_{4-x}$Se$_4$. To illustrate whether the ground state is a spin liquid, a good way is to combine the measurements of low temperature specific heat and thermal conductivity. However, since the present sample is a polycrystalline one, the transport of any kind of supermagnons or quantum spin fluctuations will be hindered by the grain boundaries or the interface between the grains. Thus it may not be conclusive even the thermal conductivity experiment does not show a residual term of $\kappa$$_0$/$T$ in low temperature region. Such experiments on single crystals of CsFe$_{4-x}$Se$_4$ are highly desired.

\begin{figure}
  \includegraphics[width=8.5cm]{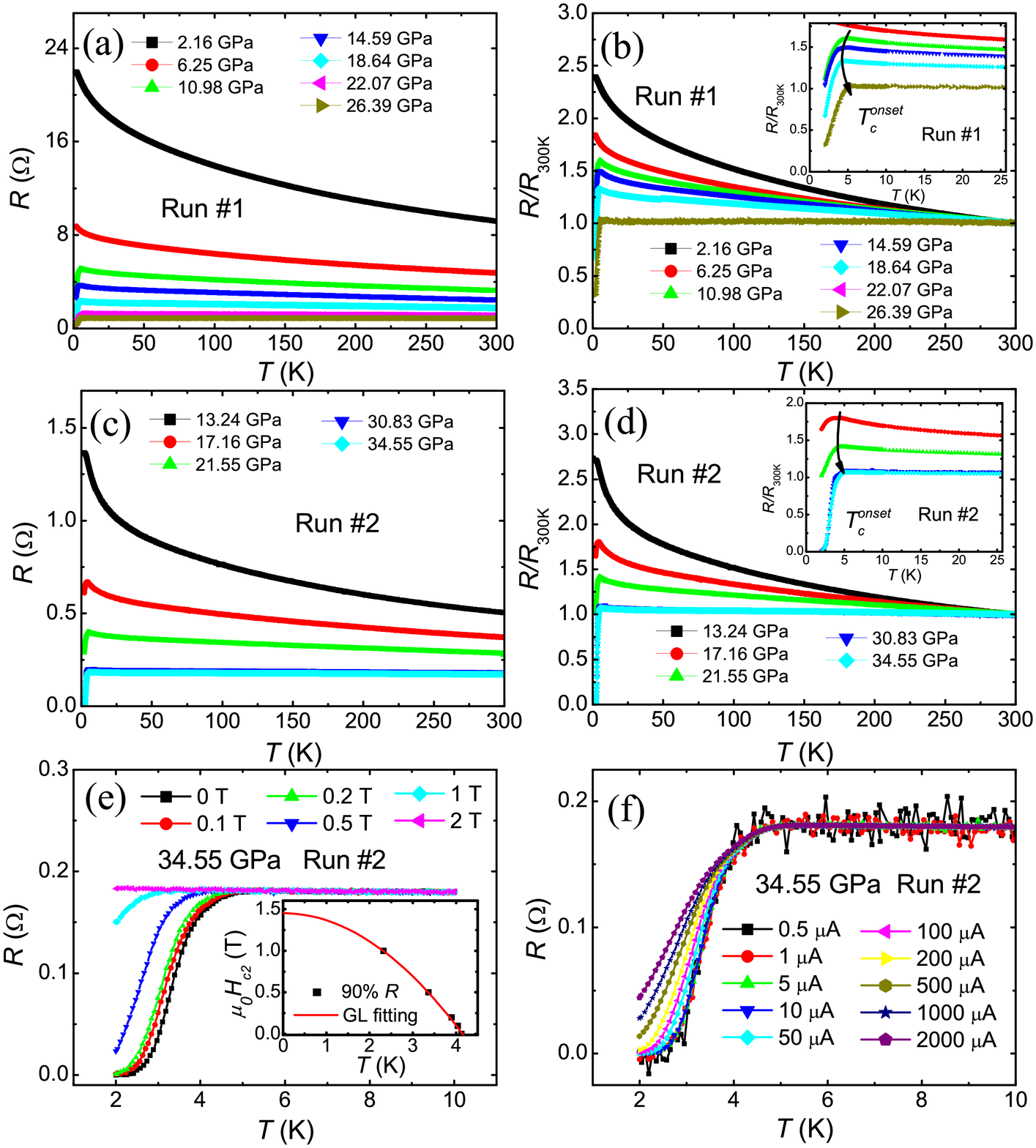}
\caption{Pressure induced superconductivity in CsFe$_{4-x}$Se$_4$. (a), (b), Temperature dependent resistance and normalized resistance at high pressures for Run $\#$1. Inset in (b) shows an enlarged view on the normalized data in low temperature region (0-25 K). (c), (d), Temperature dependent resistance and normalized resistance at high pressures for Run $\#$2. Inset in (d) shows an enlarged view of the same data in low temperature region (0-25 K). (e), Superconducting transition curves measured under different magnetic fields at 34.55 GPa of Run $\#$2. Inset in (e) shows phase diagram of the upper critical field. (f), Superconductivity measured with different electric currents at 34.55 GPa of Run $\#$2.} \label{fig4}
\end{figure}

As an effective and clean way to manipulate the properties of a compound, high pressure is also applied to this material. The results are shown in Fig.~\ref{fig4}(a)-(f). For the high pressure measurements, we used different diamond anvil cells (DAC) for different runs of experiments. The culet size of DAC for Run $\#$1 (Fig.~\ref{fig4}(a),(b)) is 400 $\mu$m while for Run $\#$2 (Fig.~\ref{fig4}(c),(d)) is 300 $\mu$m, so that the maximum of pressure applied in Run $\#$2 is higher than that in Run $\#$1. By increasing the applied pressure, the insulator-like behavior can be successively suppressed and the system undergoes a transition from an insulator to a bad metal. The resistance decreases monotonously with applied pressure, meanwhile the residual-resistivity-ratio ($RRR$, $R$$_{300K}$/$R$$_{6K}$) increases from 0.42 to 0.96 in Run $\#$1, and from 0.39 to 0.93 in Run $\#$2. Accompanying the emergence of metallicity, superconductivity gradually appears. The resistivity starts to drop at about 5.1 K under the pressure of 10.98 GPa in Run $\#$1 and 13.24 GPa in Run $\#$2. As pressure keeps increasing, the $T$$_c^{onset}$ changes slightly and the ratio of superconducting phase becomes larger until zero resistance is measured at 30.83 GPa in Run $\#$2. The enlarged views of the data in low temperature region (0-25 K) are shown in the inset of Fig.~\ref{fig4}(b) and (d), respectively. The low temperature resistance data under different magnetic fields up to 2 T at 34.55 GPa are also obtained and shown in Fig.~\ref{fig4}(e). One can see that the magnetic field can easily suppress superconductivity. By taking 90$\%$ of the normal state resistance as the criterion to determine $T$$_c$, we can get the phase diagram of the upper critical field, which is shown in the inset of Fig.~\ref{fig4}(e). By fitting the data with the formula based on the Ginzburg-Landau theory $H$$_{c2}$($T$)=$H$$_{c2}$(0)$\cdot$[1-($T$$_c$/$T$$_0$)$^2$], the $H$$_{c2}$(0) is about 1.45 T, which is relatively small compared with that of other FeSe based superconductors.

One may argue that the superconductivity observed here may arise from some impurity phases. However, based on the obtained message of $T$$_c$ and $H$$_{c2}$(0), we can prove that superconductivity observed here is induced by the main phase CsFe$_{4-x}$Se$_4$, rather than other impurity phases. Judging from the values of $T$$_c$, we can exclude the possibility that the superconducting phase comes from possible impurities like Fe\cite{26}, Cs$_{0.8}$Fe$_{2}$Se$_{2}$\cite{27} or FeSe\cite{43,3} since their $T$$_c$ values are very different from the observed 5.1 K here. To corroborate this, we emphasize that no clear trace of these impurity phases are visible from the XRD data. There is a concern that the superconductivity may come from the Se element, whose $T$$_c$ at high pressure is about 6.5 K. We then measure the resistivity of elemental Se by applying pressures, the data are shown in Supplementary Figure 3. Comparing with Se at the pressure of about 30 GPa, the $T$$_c$ of CsFe$_{4-x}$Se$_4$ is lower and the phase lines of $H$$_{c2}$($T$) are also different. Generally, the upper critical field $H$$_{c2}$($T$) of the pressurized CsFe$_{4-x}$Se$_4$ is much lower than that of Se and no any trace of Se element can be observed from the XRD data. Thus this possibility can also be ruled out. In addition, we want to emphasize that the appearance and vanishing of superconductivity in present system follows very well in processes with increasing and descending pressure. After superconductivity appears at a high pressure, when we lower down the pressure, the superconductivity gradually disappears and the strong insulating feature shows up again. The data with decreasing pressure are shown in Supplementary Figure 4. All these indicate that the superconductivity here is an intrinsic property of CsFe$_{4-x}$Se$_4$ under pressure. With these arguments and several rounds of control experiments, we can safely conclude that the superconductivity observed here is induced in the CsFe$_{4-x}$Se$_4$ system by pressure effect.

The absence of zero resistance in Run $\#$1 and Run $\#$3 may be because we have used a relatively large measuring current (1000 $\mu$A) concerning the very small sample size of the DAC. In Fig.~\ref{fig4}(f), we show the superconducting transitions under different applied currents. Interestingly, we find that the superconducting transition can be affected easily by the applied current, although the onset temperature does not shift with the current. With a small measuring current, we can see the zero resistance state, however the zero resistance state is lost when the measuring current becomes large. We must emphasize that the vanishing of the zero resistance state is not due to heating effect, since the onset transition temperature does not change at all with different currents. We would argue that this mimics the picture that the superconducting state is formed by the diluted superfluid density on top of the background of a correlated bad metal. The superconducting phenomenon can be repeatedly observed in each run of our high pressure measurements. Since our sample is polycrystalline in nature, the effect of grain boundaries cannot be ignored. However, under such a high pressure, the grain boundary issue on the critical current density may not be serious. This happens for the following reasons. Firstly, according to Pippard relation $\xi$=$\hbar$$v$$_F$/$\Delta$, the low $T$$_c$ may correspond to a small gap and large coherence length, and large coherence length will overcome the problem of grain boundaries. Secondly, the CsFe$_{4-x}$Se$_4$ system may not be that anisotropy like many cuprates, and under a high pressure, the weak-link effect is weakened, this further weakens the influence of the grain boundaries on critical current density.

\section{Discussion}

Up to now, we have found several interesting features of the new compound CsFe$_{4-x}$Se$_4$. Firstly, the material shows a strong insulator-like behavior which cannot be described by the model of a band insulator, while the temperature dependence of resistivity can be fitted to the 3D VRH model with correlations\cite{16} in wide temperature region. The DFT calculation shows that this compound should be a metal with the 3$d$$_{xz/yz/xy}$ orbits as the dominant ones\cite{15}. In the original report, the authors attribute this insulator-like behavior to the Fe vacancy\cite{15}. This is unlikely since the Fe vacancy is only about 1.25$\%$ in our samples according to our EDS analysis. Furthermore, given the presence of certain amount of Fe vacancies, it is hard to understand why the originally metallic background would show such a strong insulator-like behavior. Secondly, we find that the specific heat shows a sizable residual coefficient $\gamma$$_0$, this is certainly unexpected for a band insulator. It is known that for a spin liquid, due to the existence of quantum spin fluctuation at zero temperature, it is expected to have a contribution for the specific heat coefficient. We thus believe that the residual linear specific heat contribution in low temperature region has a nontrivial origin, it is most likely to be related to the correlation effect in this compound. Thirdly we have observed superconductivity by suppressing this insulator-like behavior after applying a high pressure. From the upper critical field determined here, we find that the $H$$_{c2}$(0) seems to be quite low. This is different from other FeSe or FeAs based superconductors in which the upper critical field is generally quite high. More interestingly, we find that the zero resistance superconducting temperature is dependent on the applied current. This reminds us that the superconducting state achieved under high pressure may be formed by the very diluted superfluid on the background of a correlated system. The same situation occurs in the underdoped cuprate superconductors in which the superfluid density is very low and the superconducting transition temperature is determined by this Bose-Einstein condensation temperature\cite{28,29}.

Actually in the iron based superconductors, the correlation strength is orbital dependent and can be quite strong, in many cases the $d$$_{xy}$ orbit shows a strong and temperature dependent correlation effect. This has been called as the orbital selective Mottness\cite{30}. Due to the complex charge-spin-lattice interactions, the correlation effect can be induced by the Hund's coupling effect with different strength on different orbits, which may lead to the insulator-like behavior\cite{31}. Recently, the quasiparticle interference technique based on the scanning tunneling microscope measurements reveals strong and orbital selective differences of quasiparticle weight $Z$ on all detectable bands over a wide energy range in bulk FeSe\cite{32}. The authors conclude that orbital selective strong correlations dominate the parent state of iron-based high-temperature superconductivity in FeSe, even the normal state looks like a metal. In the FeSe based family, there exist some other systems, such as the BaFe$_2$Se$_3$, which has been regarded as the orbital selective Mott insulator\cite{33}. Inelastic neutron scattering results show that this system has a block antiferromagnetic state with rather large magnetic moments\cite{34}. It also tells how the orbital degrees of freedom in iron-based compounds can help to stabilize an exotic magnetic state. By applying pressure, this Mott insulator turns gradually from a strong insulator-like behavior to a metallic state and finally superconductivity is observed\cite{35}. Although the structures of BaFe$_2$Se$_3$ and the present system CsFe$_{4-x}$Se$_4$ are different, the fundamental physics may be similar, namely the orbital selective Mottness can be suppressed by pressure, charges then get more freedom to move and superconductivity is finally achieved. Superconductivity was also induced in another Fe-based compound FePSe$_3$ by pressure\cite{36}. This material has a honeycomb structure at ambient pressure and undergoes a structural transition at a high pressure. The authors attribute the emergence of superconductivity to the crossover of the spin states via the structural transition. An intuitive picture called Gossamer state was invented by Laughlin\cite{37} and developed by Zhang et al.\cite{38} to describe the unique superconducting state in cuprates, which postulates a diluted superfluid density on the correlated background. We think this picture may also be applied to our present system. Our present study illustrates that the CsFe$_{4-x}$Se$_4$ system may provide a strong correlated electronic platform for inducing unconventional superconductivity. It calls for theoretical efforts beyond the DFT calculations to understand how strong the orbital selective correlations are and how the Mottness is established in this particular system, and finally why superconductivity is induced by applying pressure. In a newly found Cu-based oxy-arsenide RE$_2$Cu$_5$As$_3$O$_2$ (RE = La, Pr, Nd)\cite{42}, it seems that the correlation effect is not very strong and a Fermi liquid behavior appears in the normal state, which is different from the present system CsFe$_{4-x}$Se$_4$. While in Fe based superconductors, most systems in the parent state show bad metal behavior depending on the subtle balance between itinerancy and localization of the $d$-orbital electrons. As far as we know, beside BaFe$_2$Se$_3$, CsFe$_{4-x}$Se$_4$ is another one which shows strong correlation effect, and superconductivity is induced by pressurizing the parent state with Mottness, which is similar to the cuprates in this regard. However, compared to BaFe$_2$Se$_3$, CsFe$_{4-x}$Se$_4$ seems to be more fundamental in structure because it possesses the basic FeSe layers.

In conclusion, we successfully synthesize the newly discovered FeSe-based compound CsFe$_{4-x}$Se$_4$ and find several intriguing physical proprieties. At ambient pressure, it is insulator-like, and the temperature dependent resistivity can be described by the 3D VRH model for correlated system. The temperature dependence of magnetization does not obey the Curie-Weiss law but shows a unique linear behavior with a negative slope versus temperature in wide temperature region. The specific heat down to 400 mK shows a sizable residual specific heat coefficient, this is unexpected for a band insulator. We intend to attribute this residual specific heat to the natural quantum spin fluctuations of the system in the zero temperature limit and the background is argued to be a Mott insulator. By applying high pressure, the insulator-like behavior is gradually suppressed, and superconductivity appears at about 5.1 K. The superconductivity can be suppressed both by external magnetic field and electric current, which suggests that the superfluid density is fragile in this system. We believe that the CsFe$_{4-x}$Se$_4$ can provide a platform to explore unconventional superconductivity established by diluted Cooper pairs on top of a strong correlation background.

\section*{ACKNOWLEDGMENTS}
We are grateful to Louis Taillefer and Shiyan Li for discussions and the preliminary measurements of low temperature thermal conductivity. This work was supported by the National Key R$\&$D Program of China (Grant No. 2016YFA0300401 and 2016YFA0401704) and National Natural Science Foundation of China (Grant No. A0402/11534005, A0402/11927809, A0402/13001167 and A0402/11674164).

\end{document}